\documentclass[runningheads]{svmult}

\usepackage{makeidx}   % allows index generation
\usepackage{graphicx}  % standard LaTeX graphics tool
\usepackage{subeqnar}  % subnumbers individual equations
\usepackage{multicol}  % used for the two-column index
\usepackage{physprbb}  % modified textarea for proceedings,
\makeindex             % used for the subject index
\begin{document}
\title*{When Supermassive Black Holes were growing: Clues from 
Deep X--ray Surveys}
\toctitle{Growing Black Holes: Clues from Deep X--ray Surveys}
\titlerunning{Clues from Deep X--ray Surveys}
\author{G\"unther Hasinger}
\authorrunning{G\"unther Hasinger}
\institute{Max-Planck-Institute for Extraterrestrial Physics, 84571 Garching, Germany}

\maketitle              % typesets the title of the contribution

\begin{abstract}

Merging the {\em Chandra} and {\em XMM--Newton} deep surveys with the previously identified {\em ROSAT} surveys a unique sample of almost 1000 AGN--1 covering five orders of magnitude in 0.5--2 keV flux limit and six orders of magnitude in survey solid angle with $\sim95\%$ completeness has been constructed. The luminosity--redshift diagram is almost homogeneously filled. AGN--1 are by far the largest contributors to the soft X--ray selected samples. Their evolution is responsible for the break in the total 0.5--2 keV source counts. The soft X--ray AGN--1 luminosity function shows a clear change of shape as a function of redshift, confirming earlier reports of luminosity--dependent density evolution for optical quasars and X--ray AGN. The space density evolution with redshift changes significantly for different luminosity classes, showing a strong positive evolution, i.e. a density increase at low redshifts up to a certain redshift and then a flattening. The redshift, at which the evolution peaks, changes considerably with X--ray luminosity, from $z\approx$0.5--0.7 for luminosities $\log L_x$=42--43 erg s$^{-1}$ to $z\approx2$ for $\log L_x$=45--46 erg s$^{-1}$. The amount of density evolution from redshift zero to the maximum space density also depends strongly on X--ray luminosity, more than a factor of 100 at high luminosities, but less than a factor of 10 for low X--ray luminosities. For the first time, a significant decline of the space density of X--ray selected AGN towards high redshift has been detected in the range $\log L_x$=42--45 erg s$^{-1}$, while at higher luminosities the survey volume at high--redshift is still too small to obtain meaningful densities. A comparison between X--ray and optical properties shows now significant evolution of the X--ray to optical spectral index for AGN--1. The constraints from the AGN luminosity function and evolution in comparison with the mass function of massive dark remnants in local galaxies indicates, that the average supermassive black hole has built up its mass through efficient accretion ($\epsilon\sim10\%$) and is likely rapidly spinning.

\end{abstract}

\section{Introduction}

In recent years the bulk of the extragalactic X--ray background in the 0.1-10
keV band has been resolved into discrete sources with the deepest 
{\em ROSAT}, {\em Chandra} and {\em XMM--Newton} observations  
\cite{has98,mus00,gia01,gia02,has01,ale03,wor04,bau04}. Optical identification programmes with 
Keck \cite{schm98,leh01,bar01a,bar03} and VLT \cite{szo04,fio03} find predominantly
unobscured AGN--1 at X--ray fluxes $S_X>10^{-14}$ erg cm$^{-2}$ s$^{-1}$, and a
mixture of unobscured and obscured AGN--2 at fluxes $10^{-14}>S_X>10^{-15.5}$
erg cm$^{-2}$ s$^{-1}$ with ever fainter and redder optical counterparts, while
at even lower X--ray fluxes a new population of star forming galaxies 
emerges \cite{hor00,ros02,ale02,hor03,nor04,bau04}. At optical magnitudes R$>$24 these surveys suffer 
from large spectroscopic incompleteness, but deep optical/NIR photometry can improve the 
identification completeness significantly, even for the faintest optical 
counterparts \cite{zhe04,mai04}. A recent review article \cite{bra05} summarizes the current status of X--ray deep surveys. 

The AGN/QSO luminosity function and its evolution with 
cosmic time are key observational quantities for understanding
the origin of and accretion history onto supermassive black holes,
which are now believed to occupy the centers of most galaxies.
X--ray surveys are practically the most efficient means of finding 
active galactic nuclei (AGNs) over a wide range of luminosity and 
redshift. Enormous efforts have been made by several groups to follow 
up X--ray sources with major optical telescopes around the globe, so
that now we have fairly complete samples of X--ray selected AGNs.
The most complete and sensitive sample was compiled recently by Hasinger, Miyaji and Schmidt \cite{has05}, concentrating on unabsorbed (type--1) AGN selected in
the soft (0.5--2 keV) X--ray band, where due to the previous {\em ROSAT} work
\cite{paper1,paper2} complete samples exist, with
sensitivity limits varying over five orders of magnitude in flux, and 
survey solid angles ranging from the whole high galactic latitude sky to 
the deepest pencil-beam fields. These samples allowed to construct 
luminosity functions over cosmological timescales, with an
unprecedented accuracy and parameter space.

\begin{table}[ht]
\caption[]{The soft X--ray sample}
\begin{center}
\begin{tabular}{@{}lccccccc@{}}
\hline
Survey$^{\rm a}$ & Solid Angle &  $S_{\rm X14,lim}$ & $N_{\rm tot}$ & $N_{\rm AGN-1}$$^{\rm b}$ & $N_{\rm unid}^{\rm c}$ \\
{}     & [deg$^2$] &[cgs]          &               &                &               &\\
\hline
RBS    & 20391     & $\approx 250$ &   901    & 203  &     0  \\
SA--N   & 684.0--36.0  & 47.4--13.0   &   380    & 134  &     5  \\    
NEPS   & 80.7--1.78   & 21.9--4.0   &   262    & 101  &     9  \\  
RIXOS  & 19.5--15.0   & 10.2--3.0    &   340    & 194  &    14  \\   
RMS    & 0.74--0.32   & 1.0--0.5     &   124    &  84  &     7  \\
RDS/XMM & 0.126--0.087 & 0.38--0.13   &  81    &  48  &     8  \\
CDF--S   & 0.087--0.023   & 0.022--0.0053& 293   & 113  &     1  \\
CDF--N   & 0.048--0.0064  & 0.030--0.0046& 195   &  67  &    21  \\
\hline
Total  &           &               &  2566    & 944  &    57  \\        
\hline
\end{tabular}
\end{center}
\label{tab:samp}

$^{\rm a}$ Abbreviations -- RBS: The {\em ROSAT} Bright Survey \cite{schw00}; SA--N: {\em ROSAT} Selected Areas North \cite{app98}; NEPS: {\em ROSAT} North Ecliptic Pole Survey \cite{gio03}; RIXOS: {\em ROSAT} International X--ray Optical Survey \cite{mas00}, RMS: {\em ROSAT} Medium Deep Survey, consisting of deep PSPC pointings at the North Ecliptic Pole \cite{bow96}, the UK Deep Survey \cite{mch98}, the Marano field \cite{zam99} and the outer parts of the Lockman Hole \cite{schm98,leh00}; RDS/XMM: {\em ROSAT} Deep Survey in the central part of the Lockman Hole, observed with {\em XMM--Newton} \cite{leh01,mai02,fad02}; CDF--S: The {\em Chandra} Deep Field South \cite{szo04,zhe04,mai04}; CDF--N: The {\em Chandra} Deep Field North \cite{bar01a,bar03}. 

$^{\rm b}$ Excluding AGNs with $z<0.015$. 

$^{\rm c}$ Objects without redshifts, but hardness ratios consistent with 
type--1 AGN. 

\end{table}

\section{The X--ray selected AGN--1 sample}

For the derivation of the X--ray luminosity function and cosmological 
evolution of AGN well--defined flux--limited 
samples of active galactic nuclei have been chosen, with flux limits and survey solid angles ranging over five and six orders of magnitude, respectively (see Table 1). To be able to utilize the massive amount of optical identification work performed previously on a large number of shallow to deep {\em ROSAT} 
surveys, the analysis was restricted to samples selected in the 0.5--2 
keV band. In addition to the {\em ROSAT} surveys already used in 
\cite{paper1,paper2}, data from the recently published 
{\em ROSAT} North Ecliptic Pole Survey (NEPS) \cite{gio03,mul04}, 
from an {\em XMM--Newton} observation of the Lockman Hole
\cite{mai02} as well as the {\em Chandra} Deep Fields South 
(CDF--S) \cite{szo04,zhe04,mai04} and North 
(CDF--N) \cite{bar01a,bar03} were included. In order to avoid 
systematic uncertainties introduced by the varying and a priori unknown
AGN absorption column densities only unabsorbed
(type--1) AGN, classified by optical and/or X--ray methods were selected.   
We are using here a definition of type--1 AGN, which is largely based on the presence of broad Balmer emission lines and small Balmer decrement in the optical spectrum of the source (optical type--1 AGN, e.g. the ID classes a, b, and partly c in \cite{schm98}, which largely overlaps the class of X--ray type--1 AGN defined by their X--ray luminosity and unabsorbed X--ray spectrum \cite{szo04}. However, as Szokoly et al show, at low 
X--ray luminosities and intermediate redshifts the optical AGN classification often breaks down because of the dilution of the AGN excess light by the stars in the host galaxy (see e.g. \cite{mor02}), so that only an X--ray classification scheme can be utilized. Schmidt et al. \cite{schm98} have already introduced the X--ray luminosity in their classification. For the deep XMM--Newton and {\em Chandra} surveys in addition the X--ray hardness ratio was used to discriminate between X--ray type--1 and type--2 AGN, following \cite{szo04}. 

Most \hbox{($\approx 70$--100\%)} of the extragalactic \hbox{X--ray} sources found in both the deep and wider \hbox{X--ray} surveys with Chandra and XMM-Newton are AGN of some type. Starburst and normal galaxies make increasing fractional contributions at the faintest \hbox{X--ray} flux levels, but even in the \hbox{CDF-N} they represent \hbox{$\sim 20$--30\%} of all sources (and create $\sim 5$\%\ of the XRB). The observed AGN sky density in the deepest \hbox{X--ray} surveys is $\approx 7200$~deg$^{-2}$, about an order of magnitude higher than that found at any other wavelength \cite{bau04}. This exceptional effectiveness at finding AGN arises because \hbox{X--ray} selection (1) has reduced absorption bias and minimal dilution by host-galaxy starlight, and (2) allows concentration of intensive optical spectroscopic follow-up upon high-probability AGN with faint optical counterparts (i.e., it is possible to 
probe further down the luminosity function). 

   \begin{figure}
   \centering
   \includegraphics[width=6cm]{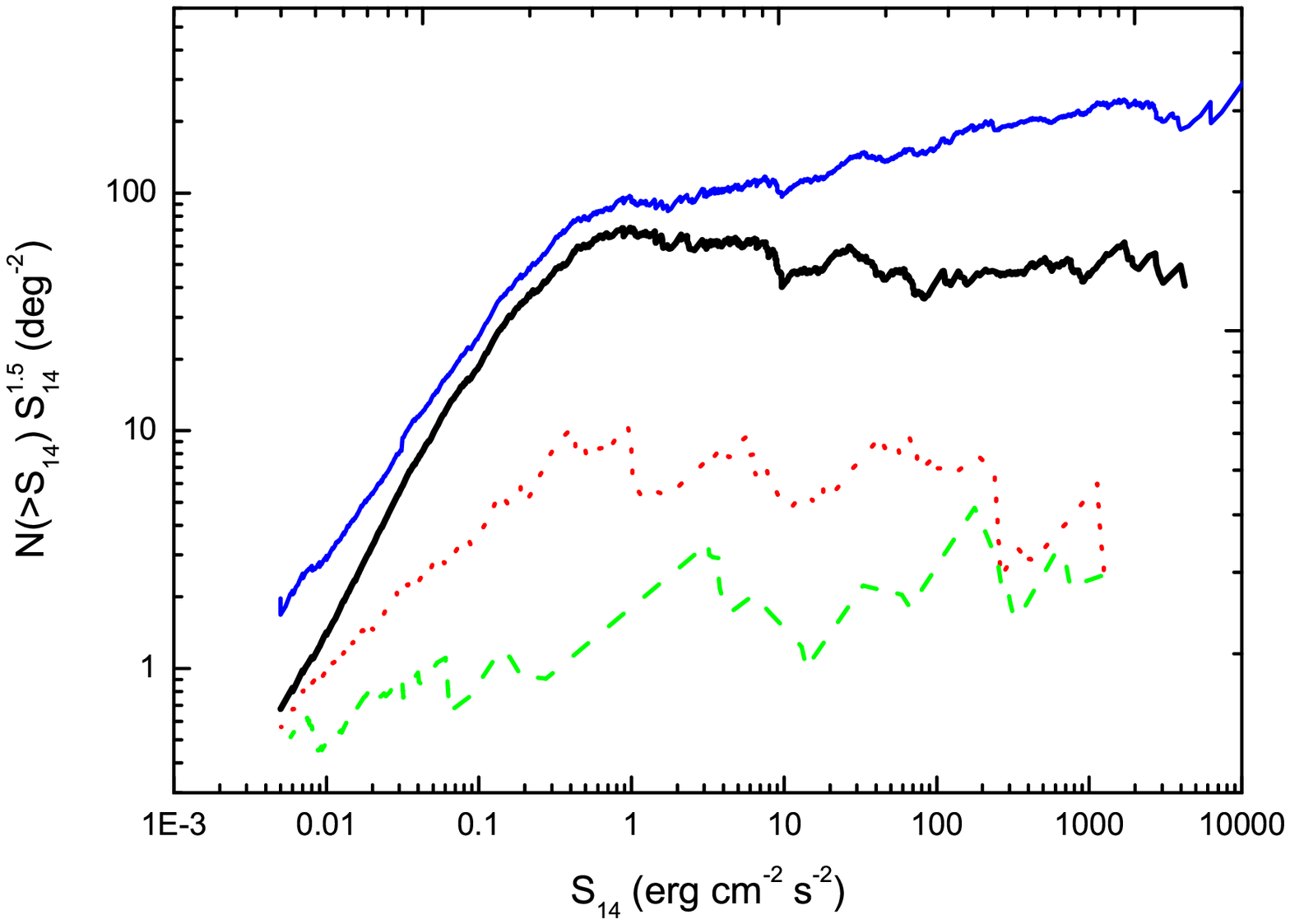}
   \includegraphics[width=6cm]{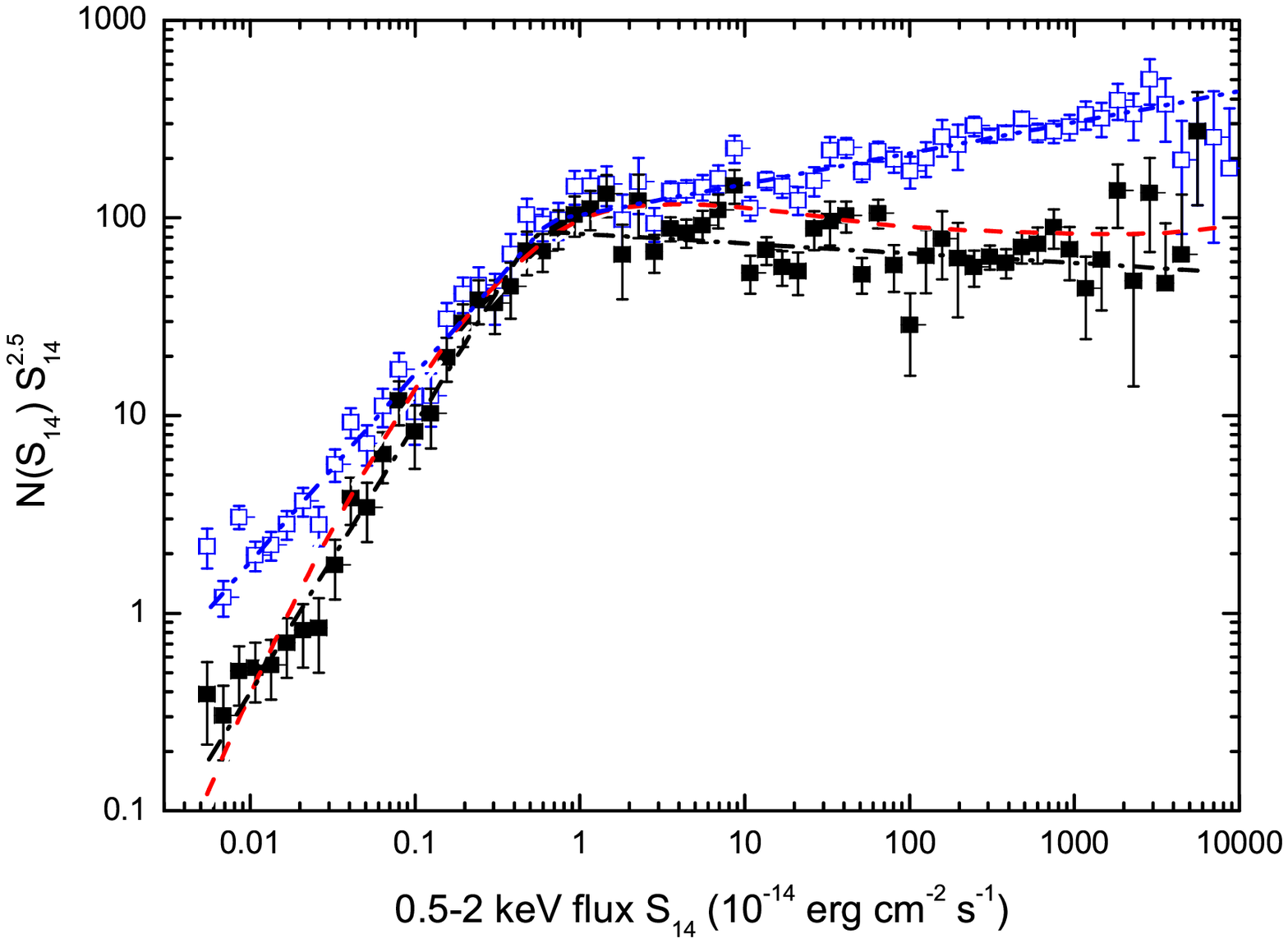}
   \caption{(a) Cumulative number counts N($>$S) for the total sample
   (upper blue thin line), the AGN--1 subsample (lower black thick line),
   the AGN--2 subsample (red dotted line) and the galaxy subsample
   (green dashed line).
(b) Differential number counts of the total sample of X--ray sources
   (open squares) and the AGN--1 subsample (filled squares).  
   The dot-dashed lines refer to broken powerlaw fits to the differential
   source counts (see text). The dashed red line shows 
   the prediction for type--1 AGN (from \cite{has05}). }
   \label{fig:ns} 
   \end{figure}

\section{Number Counts and Resolved Background Fraction}

Based on deep surveys with Chandra and XMM-Newton, the \hbox{X--ray} log(N)--log(S) relation has now been determined down to fluxes of $2.4\times10^{-17}$ erg cm$^{-2}$ s$^{-1}$, $2.1\times10^{-16}$ erg cm$^{-2}$ s$^{-1}$, and $1.2\times10^{-15}$ erg cm$^{-2}$ s$^{-1}$ in the 0.5--2, 2-10 and 5-10 keV band, respectively \cite{bra01b,has01,ros02,mor03,bau04}. Figure~\ref{fig:ns}a shows the normalized cumulative source counts  $N(>S_{\rm X14}) S_{\rm X14}^{1.5}$.
The total differential source counts, normalized to a Euclidean behaviour (dN/d$S_{\rm X14} \times S_{\rm X14}^{2.5}$) is shown with open symbols in Figure~\ref{fig:ns}b. 
Euclidean source counts would correspond to horizontal lines in these graphs. For the total source counts, the well-known broken powerlaw behaviour is confirmed with high precision. A broken power law fitted to the differential source counts yields power law indices of $\alpha_b=2.34\pm0.01$ and $\alpha_f=1.55\pm0.04$ for the bright and faint end, respectively, a break flux of $S_{\rm X14}= 0.65\pm0.10$ and a normalisation of  dN/d$S_{\rm X14} = 103.5 \pm 5.3$ deg$^{-2}$ at $S_{\rm X14} = 1.0$ with a reduced $\chi^2$=1.51. 
We see that the total source counts at bright fluxes, as determined by the {\em ROSAT} All-Sky Survey data, are significantly flatter than Euclidean, consistent with the discussion in \cite{has93}. Moretti et al. \cite{mor03}, on the other hand, have derived a significantly steeper bright flux slope ($\alpha_b \approx 2.8$) from {\em ROSAT} HRI pointed observations. This discrepancy can probably be attributed to the selection bias against bright sources, when using pointed observations where the target area has to be excised.

The ROSAT HRI Ultradeep Survey had already resolved 70-80\% of the extragalactic 0.5--2 keV XRB into discrete sources, the major uncertainty being in the absolute flux level of the XRB. The deep Chandra and XMM-Newton surveys have now increased the resolved fraction to 85-100\% \cite{mor03,wor04}. Above 2 keV the situation is complicated on one hand by the fact, that the HEAO-1 background spectrum \cite{mar80}, used as a reference over many years, has a $\sim 30\%$ lower normalization than several earlier and later background measurements (see e.g. \cite{mor03}). Recent determinations of the background spectrum with XMM-Newton \cite{mol04} and RXTE \cite{rev03} strengthen the consensus for a 30\% higher normalization, indicating that the resolved fractions above 2 keV have to be scaled down correspondingly. On the other hand, the 2-10 keV band has a large sensitivity gradient across the band. A more detailed investigation, dividing the recent 770 ksec XMM-Newton observation of the Lockman Hole into finer energy bins, comes to the conclusion, that the resolved fraction decreases substantially with energy, from over 90\% below 2 keV to less than 50\% above 5 keV \cite{wor04}.

Type--1 AGN are the most abundant population of soft X--ray sources. For the determination of the AGN--1 number counts we include those unidentified sources, which have hardness ratios consistent with AGN--1 (a contribution of $\sim 6\%$, see Table~\ref{tab:samp}). Figure~\ref{fig:ns} shows, that the break in the total source counts at intermediate fluxes is produced by type--1 AGN, which are the dominant population there. Both at bright fluxes and at the faintest fluxes, type--1 AGN contribute about 30\% of the X--ray source population. At bright fluxes, they have to share with clusters, stars and BL-Lac objects, at faint fluxes they compete with type--2 AGN and normal galaxies (see Fig. \ref{fig:ns}a and \cite{bau04}). A broken power law fitted to the differential AGN--1 source counts yields power law indices of $\alpha_b=2.55\pm0.02$ and $\alpha_f=1.15\pm0.05$ for the bright and faint end, respectively, a break flux of $S_{\rm X14}= 0.53\pm0.05$, consistent with that of the total source counts within errors, and a normalisation of of dN/d$S_{\rm X14} = 83.2 \pm 5.5$ deg$^{-2}$ at $S_{\rm X14} = 1.0$ with a reduced $\chi^2$=1.26. The AGN--1 differential source counts, normalized to a Euclidean behaviour (dN/d$S_{\rm X14} \times S_{\rm X14}^{2.5}$) is shown with filled symbols in Figure~\ref{fig:ns}.

\section{The Soft X--ray Luminosity Function and Space Density Evolution}
\label{sec:exlf}

\begin{figure*}
\centering
\includegraphics[width=12cm]{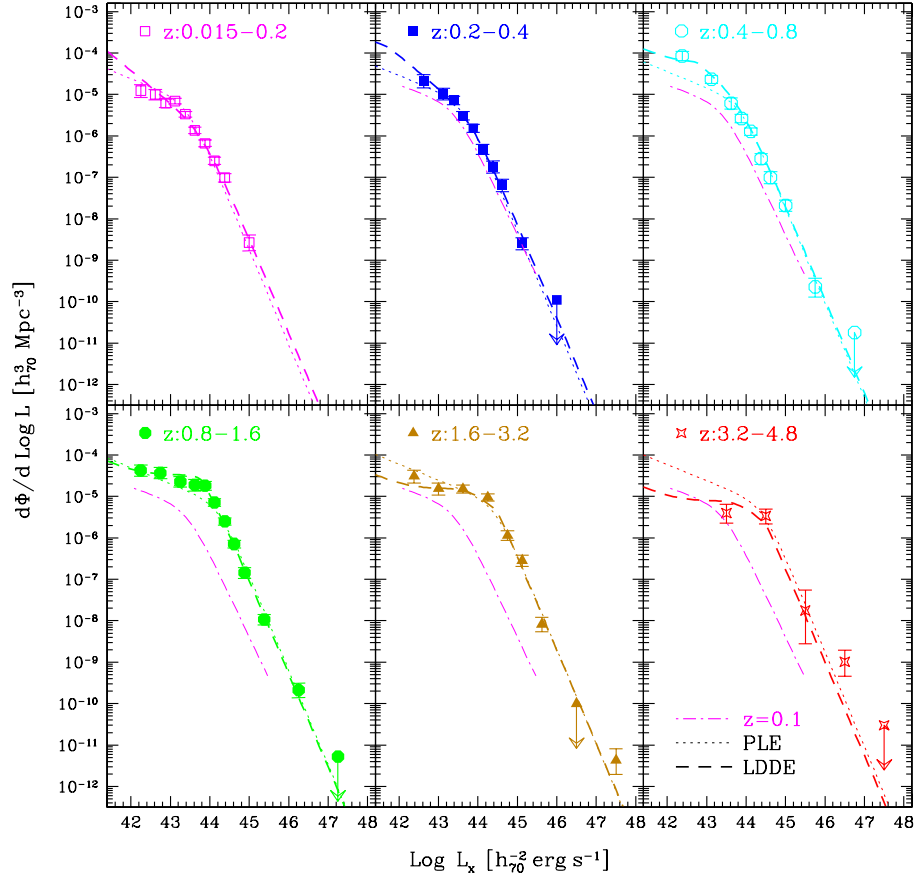}
\caption{The soft X--ray luminosity function
 of the type--1 AGN sample in different redshift shells for the
 nominal case as labelled. The error bars correspond to 68\% Poisson 
 errors of the number of AGNs in the bin. The best--fit two power-law
 model for the $0.015<z<0.2$ shell are overplotted in the higher
 redshift panels for reference. The dotted and dashed lines give
 the best--fit PLE and LDDE models (from \cite{has05}).}
\label{fig:exlf}
\end{figure*}

Hasinger, Miyaji and Schmidt \cite{has05} have employed two different methods
to derive the AGN--1 X--ray luminosity function and its
evolution. The first method uses a variant of the $1 \over V_a$ method, which was developed in \cite{paper1}. The binned luminosity
function in a given redshift bin $z_i$ is derived by dividing the observed 
number $N_{obs}(L_x,z_i)$ by the volume appropriate to the redshift range 
and the survey X--ray flux limits and solid angles. To evaluate the bias in 
this value caused by a gradient of the luminosity function across the bin, 
each of the luminosity functions is fitted by an analytical function. 
This function is then used to predict $N_{mdl}(L_x,z_i)$. Correcting the 
luminosity function by the ratio $N_{obs}/N_{mdl}$ takes care of the 
bias to first order. 

The second method uses unbinned data. Individual $V_{\rm max}$
of the RBS sources are used to evaluate the zero-redshift luminosity function. This is free of
the bias described above: using this luminosity function to derive the
number of expected RBS sources matches the observed numbers precisely.
In the subsequent derivation of the evolution, i.e., the space density
as a function of redshift, binning in
luminosity and redshift is introduced to allow evaluation of the results.
Bias at this stage is avoided by iterating the parameters of an analytical 
representation of the space density function. Together with the zero-redshift
luminsity function this is used to predict $N_{mod}(L_x,z_i)$ for the
surveys. The observed densities in the bins are derived by multiplying the
space density value by the ratio $N_{obs}(L_x,z_i)/N_{mod}(L_x,z_i)$.
At this stage, none of the densities are derived by dividing a number by
a volume.

The other difference between the two methods is in the treatment of missing
redshifts for optically faint objects. In the binned method, all AGN
without redshift with $R > 24.0$ were assigned the central redshift of
each redshift bin to derive an upper boundary to the luminosity function.
In the unbinned method, the optical magnitudes of the RBS sources were
used to derive the optical redshift limit corresponding to $R = 24.0$.
The $V_{\rm max}$ values for surveys (such as CDF--N) spectroscopically
incomplete beyond $R = 24.0$ were based on the smaller of the X--ray and
optical redshift limits.

\begin{figure}
  \centering

  \includegraphics[width=12cm]{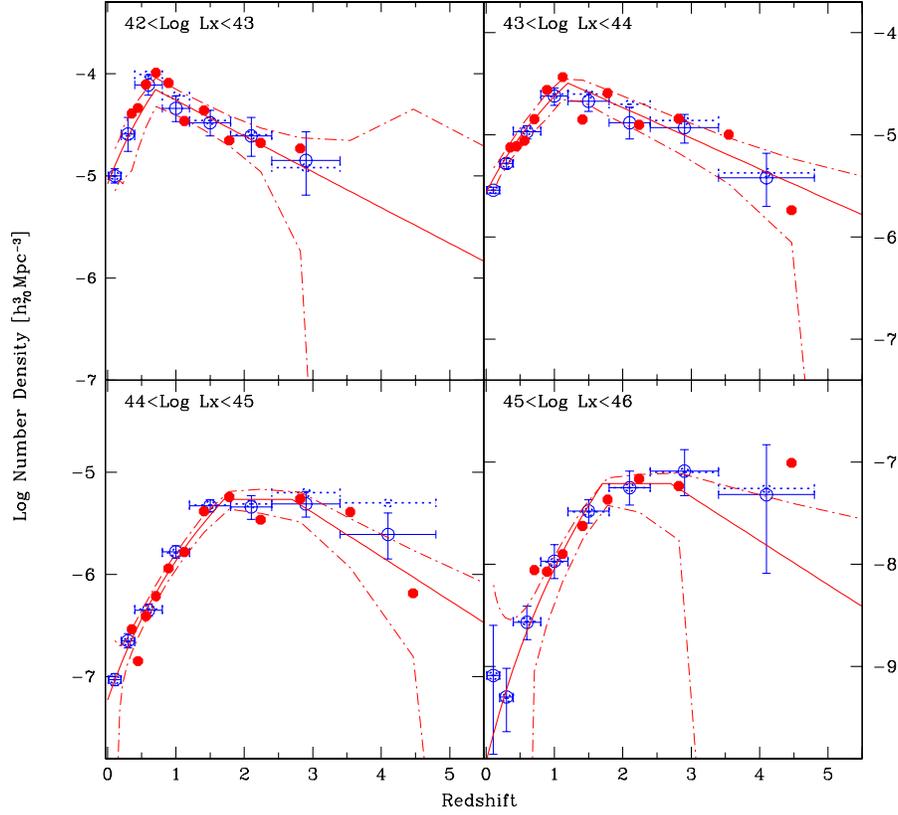}
  \caption{Comparison between the space densities derived  
           with two different methods. The blue datapoints with error 
           bars refer to the binned treatment using the 
           N$_{\rm obs}$/N$_{\rm mdl}$ method, the dashed horizontal 
           lines corresponding to the maximum contribution of unidentified
           sources. The thin and thick red lines and dots refer to the 
           unbinned method (from \cite{has05}).}
  \label{fig:comp_ma_ta}
\end{figure}

Figure \ref{fig:exlf} shows the luminosity function derived this way in different redshift shells. A change of shape of the luminosity function with redshift is clearly seen and can thus rule out simple density or luminosity evolution models. In a second step, instead of binning into redshift shells, the sample has been cut into different luminosity classes and the evolution of the space density with redshift was computed. Figure~\ref{fig:comp_ma_ta} shows a direct comparison between the binned and unbinned determinations of the space density, which agree very well within statistical errors. 

The fundamental result is, that the space density of lower--luminosity AGN--1 peaks at significantly lower redshift than that of the higher--luminosity (QSO--type) AGN. Also, the amount of evolution from redshift zero to the peak is much less for lower--luminosity AGN. The result is consistent with previous determinations based on less sensitive and/or complete data, but for the first time our analysis shows a high-redshift decline for all luminosities $L_X < 10^{45}$~erg~s$^{-1}$ (at higher luminosities the statistics is still inconclusive).  
Albeit the different approaches and the still existing uncertainties, it is very reassuring that the general properties and absolute values of the space density are very similar in the two different derivations in. 

A luminosity-dependent density evolution (LDDE) model has been fit to the
data. Even though the sample is limited to soft X--ray-selected type--1 AGN,
the parameter values of the overall LDDE model are surprisingly close to those obtained by Ueda et al. 2003 for the intrinsic (de-absorbed) luminosity function of hard X--ray selected obscured and unobscured AGN, except for the normalization, where Ueda et al. reported a value about five times higher. 

These new results paint a dramatically different 
evolutionary picture for low--luminosity AGN 
compared to the high--luminosity QSOs. While 
the rare, high--luminosity objects can form and 
feed very efficiently rather early in the 
Universe, with their space density declining
more than two orders of magnitude at redshifts below z=2,
the bulk of the AGN has to wait much 
longer to grow with a decline of space density
by less than a factor of 10 below a redshift of 
one. 
The late evolution 
of the low--luminosity Seyfert population is very 
similar to that which is required to fit the 
Mid--infrared source counts and background 
\cite{fra02} and 
also the bulk of the star formation in the 
Universe \cite{mad98}, while the 
rapid evolution of powerful 
QSOs traces more the merging history of 
spheroid formation \cite{fra99}.

This kind of anti--hierarchical Black Hole growth 
scenario is not predicted in most of the semi--analytic
models based on Cold Dark Matter structure formation
models (e.g. \cite{kau00,wyi03}). This could indicate two modes 
of accretion and black hole growth with 
radically different accretion efficiency 
(see e.g. \cite{dus02}). A
self--consistent model of the black hole growth which
can simultaneously explain the anti--hierarchical 
X--ray space density evolution and the local black 
hole mass function derived from the $M_{BH}-\sigma$
relation assuming two radically different modes of
accretion has recently been presented in \cite{mer04}.

\section{Optical versus X--ray selection of AGN--1}

 \begin{figure}
 \includegraphics[width=12cm]{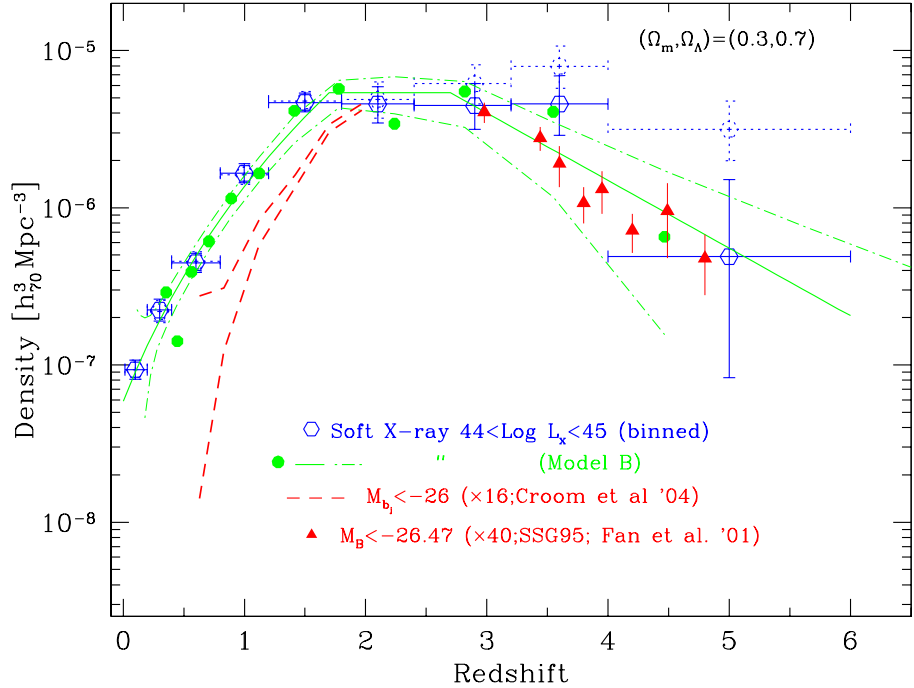}
  \caption{Comparison of the space density of luminous QSOs
    between optically selected and soft X--ray selected samples
    (from \cite{has05}).
    The X--ray number densities are plotted for the luminosity 
    class ${\rm log\,}L_{\rm x}=44-45$, both for the 
    binned and unbinned analysis with the same symbols 
    as in Fig.~\ref{fig:comp_ma_ta}.
    The dashed lines represent the one sigma range for 
    $M_{b_{r,j}}<-26.0$ from \cite{cro04},
    multiplied by a factor of 16 to match the X--ray 
    space density at z=2. 
    The triangles at $z>2.7$ with  1$\sigma$ errors are from 
    \cite{ssg95} (SSG95) 
    and \cite{fan01} after a cosmology conversion (see text) 
    and a scaling by a factor of 40 to match with the soft X--ray
    density at $z\sim 2.7$. As discussed in this paper, both the rise and
    the decline of the space density, behavior changes with $L_x$ and 
    therefore that the comparison can only be illustrative.}
 \label{fig:zevcomp}
\end{figure}

The space density of soft X--ray selected QSOs from the Hasinger et al. sample is compared to the one of optically-selected QSOs at the most luminous end in Fig.~\ref{fig:zevcomp}. The $z<2$ number density curve for optically selected QSOs ($M_{b_{\rm J}}<-26.0$) is from the combination of the 2dF and 6dF QSO redshift surveys \cite{cro04}. The $z>2.7$ number densities from \cite{ssg95}
and \cite{fan01} have been originally given for $H_0$=50 km s$^{-1}$ Mpc$^{-1}, \Omega_m=1, \Omega_\Lambda=0$. Their data points have been converted to $H_0$=70 km s$^{-1}$ Mpc$^{-1}, \Omega_m=0.3, \Omega_\Lambda=0.7$ and the $M_{\rm B}$ threshold has been re-calculated with an assumed spectral index of $\alpha_o=-0.79$ ($f_\nu\propto\nu^{\alpha_o}$), following e.g. \cite{vig03}. The plotted curve from \cite{ssg95,fan01} 
is for $M_B<-26.47$ under these assumptions. A small correction of densities due to the cosmology conversion causing redshift-dependent luminosity thresholds 
has also been made, assuming $d\Phi/dlogL_B\propto L_B^{-1.6}$ \cite{fan01}. The space density for the soft X--ray QSOs for the luminosity class $44<logL_x<45$ has been plotted both for the binned and unbinned determination. The Croom et al. \cite{cro04} space density had to be scaled up by a factor of 16 in order to match the X--ray density at $z\sim 2$. The Schmidt, Schneider \& Gunn / Fan et al. data points have been scaled by a factor of 40 to match the soft 
X--ray data at $z=2.7$ in the plot. There is relatively little 
difference in the density functions between the X--ray and optical
QSO samples, although we have to keep in mind, that both the rise 
and the decline of the space density is varying with X--ray luminosity,
so that this comparison can only be illustrative until larger 
samples of high--redshift X--ray selected QSOs will be available.  

 \begin{figure}
 \centering
 \includegraphics[width=12cm]{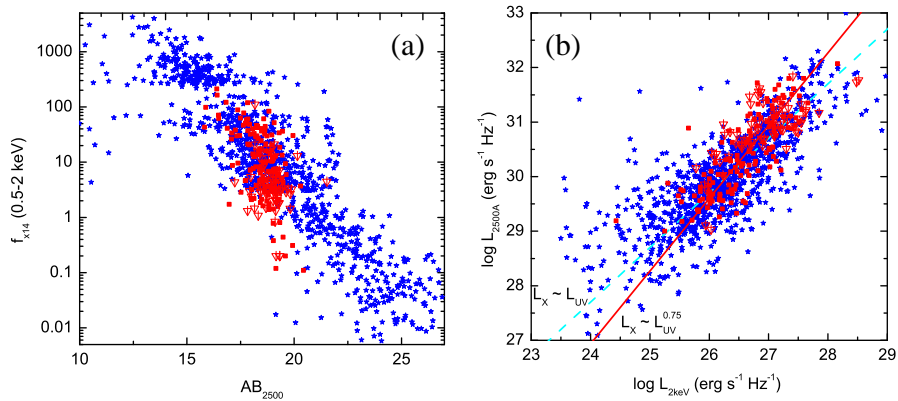}
  \caption{(a): comparison of X--ray fluxes and AB$_{2500}$ UV magnitudes
  for the sample of $\sim 1000$ X--ray selected type-1 AGN from \cite{has05} (blue points) with the ROSAT-observed optically selected SDSS QSOs \cite{vig03}. Filled red squares give the standard SDSS QSO, while open red squares give the specifically selected high-redshift SDSS sample (see \cite{vig03}). (b): Monochromatic 2 keV X--ray versus 2500 $\AA$ ~UV luminosity for the same samples. The blue (dark) solid line shows a linear relation between the two luminosities, while the yellow (light) solid line gives the non-linear relation $L_X \sim L_{UV}^{0.75}$ from the literature \cite{vig03}. }
\label{fig:x_uv}
\end{figure}

As a next step we directly study the X--ray and optical fluxes and luminosities of our sample objects and compare this with the optically selected QSO sample of Vignali et al. \cite{vig03} based on SDSS-selected AGN serendipitously observed in ROSAT PSPC pointings. Because of the inhomogeneous nature and different systematics of the different surveys entering our sample, the optical/UV magnitudes of our objects have unfortunately much larger random and systematic errors and are based on fewer colours than the excellent SDSS photometry. In our preliminary analysis we therefore calculated the AB$_{2500}$ magnitudes simply extrapolating or interpolating the observed magnitudes in the optical filters closest to the redshifted 2500 $\AA$ band, assuming an optical continuum with a power law index of -0.7, i.e. not utilizing the more complicated QSO spectral templates including emission lines which have been used in \cite{vig03}. A spectroscopic correction for the host galaxy contamination, as done for the SDSS sample, was also not possible for our sample, however, for a flux and redshift-selected sample of 94 RBS Seyferts \cite{salvato} we have morphological determinations of the nuclear versus host magnitudes (see below). In all other aspects of the analysis we follow the Vignali et al. treatment. Figure \ref{fig:x_uv}a shows 0.5--2 keV X--ray fluxes versus AB$_{2500}$ magnitudes for our sample objects (blue stars) in comparison with the Vignali et al. SDSS sample (X--ray detections are shown as filled red squares, upper limits as down--pointing triangles). It is obvious, that our multi-cone survey sample covers a much wider range in X--ray and optical fluxes than the magnitude-limited SDSS sample. Unlike the SDSS sample, our sample shows a very clear correlation between X--ray and optical fluxes, but also a wider scatter in this correlation.  

Figure \ref{fig:x_uv}b shows the monochromatic X--ray versus UV luminosity for the same data. Now the X--ray and optically selected samples cover a similar parameter range at the high luminosity end, but the X--ray selected data reach significantly lower X--ray and UV luminosities than the optically selected sample. Again, there is a larger scatter in the X--ray selected sample. The figure also shows two analytic relations between X--ray and UV luminosity: a linear relation L$_X \propto $L$_{UV}$ and the non--linear behaviour  L$_X \propto $L$_{UV}^{0.75}$ found in the literature (e.g. \cite{vig03}). While the Vignali et al. optically selected sample clearly prefers the non-linear dependence (see also \cite{bra04}), this is not true for the X-ray selected sample, which is consistent with a linear relation, apart from the behaviour at low luminosities, where significant contamination from the host galaxy is expected.  

 \begin{figure}
 \centering
 \includegraphics[width=12cm]{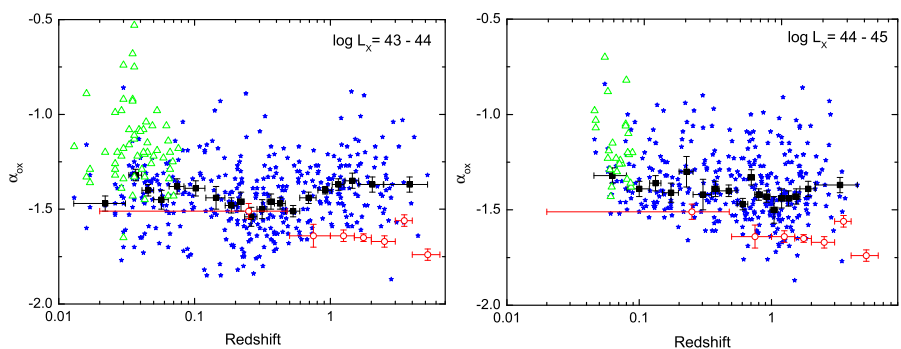}
  \caption{X--ray to optical spectral index $\alpha_{ox}$ as a function of 
    redshift for different luminosity classes for the Hasinger et al. sample 
    of $\sim1000$ soft X--ray selected type-1 AGN. Blue (dark) stars
    give the values derived using the total integrated optical light, 
    while the green triangles give the values using only the nuclear 
    component from 93 RBS AGN, derived by Salvato \cite{salvato} from 
    detailed imaging decomposition. The solid black squares with error bars
    give the median and variance of the X-ray selected sample. Open red 
    squares with error bars show the average $\alpha_{ox}$ values for
    the Vignali et al. optically selected QSOs.}
 \label{fig:z_aox}
\end{figure}

To check on any evolution of the optical to X--ray spectral index with redshift we calculated $\alpha_{ox}$ values, following \cite{vig03} for all our 
sample objects. In order to see possible luminosity--dependent evolution effectssimilar to those observed in the space density evolution, we divided our sample into the same luminosity classes as in Section \ref{sec:exlf}. Figure \ref{fig:z_aox} shows the $\alpha_{ox}$ values determined for objects in the luminosity class 43--44 and 44--45, respectively, as a function of redshift. Apart from a few wiggles, which are likely due to the omission of the emission lines in the optical AGN continuum, there is no significant evolution with redshift. The optically selected sample, on the other hand, shows a significant trend with redshift and average values inconsistent with the X-ray selected sample for most of the redshift range. The diagram also shows, that this discrepancy is likely not caused by the missing host galaxy contamination correction in our analysis. From the relatively small number of nearby (z$<$0.1) of RBS sources, where a morphological fitting procedure has been used to subtract the host emission from the total magnitude \cite{salvato} we can estimate the host dilution effect, which is clearly larger at lower X--ray luminosities and makes the discrepancy even larger. The immediate conclusion is, that the average optical to X-ray sample properties are dependent on systematic sample selection effects.    

\section{ \hbox{X--ray} Constraints on the Growth of SMBH}

The AGN luminosity function can be used to determine the masses of remnant black holes in galactic centers, using So\l tan's continuity equation argument \cite{sol82} and assuming a mass-to-energy conversion efficiency $\epsilon$. For a non-rotating Schwarzschild BH, $\epsilon$ is expected to be 0.054, while for a maximally rotating Kerr BH, $\epsilon$ can be as high as 0.37 \cite{tho74}. The AGN demography predicted, that most normal galaxies contain supermassive black holes (BH) in their centers, which is now widely accepted (e.g. \cite{kor01} and references therein). Recent determinations of the accreted mass from the optical QSO luminosity function are around $2 \epsilon_{0.1}^{-1} \cdot 10^5 M_\odot Mpc^{-3}$ \cite{cho92,yut02}. Estimates from the X--ray background spectrum, including obscured accretion power obtain even larger values: 6-9 \cite{fab99} or 8-17 \cite{elv02} in the above units, and values derived from the infrared band \cite{hae01} or multiwavelength observations \cite{bar01b} are similarly high (8-9). Probably the most reliable recent determination comes from an integration of the X--ray luminosity function. Using the Ueda et al. \cite{ued03} hard X--ray luminosity function including a correction for Compton--thick AGN normalized to the X--ray background, as well as an updated bolometric correction ignoring the IR dust emission, Marconi et al \cite{mar04} derived $\rho_{accr} \sim 3.5 \epsilon_{0.1}^{-1} \cdot 10^5 M_\odot Mpc^{-3}$. 
    
The BH masses measured in local galaxies are tightly correlated to the galactic velocity dispersion \cite{fer00,geb00}, and less tightly to the mass and luminosity of the host galaxy bulge (however, see \cite{mar03}). Using these correlations and galaxy luminosity (or velocity) functions, the total remnant black hole mass density in galactic bulges can be estimated. Scaled to the same assumption for the Hubble constant (H$_0$=70 km s$^{-1}$ Mpc$^{-1}$), recent papers arrive at different values, mainly depending on assumptions about the intrinsic scatter in the BH--galaxy correlations: $\rho_{BH}$ =  ($2.4\pm0.8$), ($2.9\pm0.5$) and $(4.6^{+1.9}_{-1.4} h_{70}^2 \cdot 10^5 M_\odot Mpc^{-3}$, respectively \cite{all02,yut02,mar04}. The local dark remnant mass function is fully consistent with the above accreted mass function, if black holes accrete with an average energy conversion efficiency of $\epsilon=0.1$ \cite{mar04}, which is the classically assumed value and lies between the Schwarzschild and the extreme Kerr solution. However, taking also into account the widespread evidence for a significant kinetic AGN luminosity in the form of jets and winds, it is predicted, that the average supermassive black hole should be rapidly spinning fast (see also \cite{elv02,yut02}). Recently, using XMM-Newton, a strong relativistic Fe K$\alpha$ line has been discovered in the average rest--frame spectra of AGN--1 and AGN--2 \cite{stre05}, which can be best fit by a rotating Kerr solution consistent with this conjecture.

{\it Acknowledgements} 
I thank my colleagues in the {\it Chandra Deep Field South} and {\it XMM--Newton Lockman Hole} projects, as well as Maarten Schmidt, Takamitsu Miyaji and Niel Brandt for the excellent cooperation on studies of the X--ray background.

\end{document}